%% file: paper_main.tex
\input{preamble}
\begin{document}
\raggedbottom
\title{Joint Bayesian Inference for Extraction of Charmonium Long Distance Matrix Elements Using PYTHIA}
\author{Yubo Liu}
\affiliation{Concord Academy, Concord, Massachusetts, 01742, United States}
\author{Rongrong Ma}
\affiliation{Brookhaven National Laboratory, Upton, New York 11973, United States}
\begin{abstract}
\input{abstract}
\end{abstract}
\maketitle
\input{introduction}

\input{methodology}
\input{results}
\input{conclusion}
\input{backmatter}
\clearpage
\onecolumngrid
\appendix
\input{spectra_appendix/appendix_spectra}
\twocolumngrid
\bibliographystyle{apsrev4-2}
\bibliography{references}
\end{document}

%% file: preamble.tex
\documentclass[aps,prd,reprint,superscriptaddress,nofootinbib,longbibliography,floatfix]{revtex4-2}
\usepackage[T1]{fontenc}
\usepackage{amsmath,amssymb,graphicx,array}
\usepackage{xurl}
\usepackage{placeins}
\usepackage{xcolor}
\providecommand{\burl}[1]{\url{#1}}
\providecommand{\jpsi}{\ensuremath{J/\psi}}
\providecommand{\psip}{\ensuremath{\psi(2S)}}

\providecommand{\me}[2]{\ensuremath{\langle\mathcal{O}^{#1}(#2)\rangle}}

\usepackage[hidelinks]{hyperref}
\makeatletter
\ifdefined\NoExcerptBibliography\else
\AtBeginDocument{\immediate\write\@auxout{\string\citation{PRDControl}}}
\fi
\makeatother
\newcounter{fitpanel}

%% file: abstract.tex
We present a joint Bayesian extraction of the nonrelativistic QCD long-distance matrix elements (LDMEs) for \(J/\psi\), \(\psi(2S)\), and \(\chi_c\) production, using short-distance coefficients and feed-down structure implemented in PYTHIA~8.317. Rather than fitting each charmonium state separately, we construct channel-resolved responses that link parent production to daughter feed-down and infer all LDMEs simultaneously from 286 prompt charmonium measurements in \(pp\) collisions at \(\sqrt{s}=5.02\)--13~TeV at the LHC. The extracted LDMEs provide a good description of the data, with an overall \(\chi^2/\mathrm{ndf}=1.18\). We also provide a positivity-constrained set that can be used directly as PYTHIA input. Using the extracted LDMEs, we predict prompt \(J/\psi\) and \(\psi(2S)\) cross sections at high transverse momentum in \(pp\) collisions at \(\sqrt{s}=200\)~GeV, which can be tested with RHIC data.

%% file: introduction.tex
\section{Introduction}\label{sec:introduction}
Charmonium states, including \(J/\psi\), \(\psi(2S)\), and \(\chi_c\), are bound states of a charm quark and an anticharm quark. Their production connects two regimes of quantum chromodynamics (QCD): the short-distance production of a charm-anticharm pair, which can be calculated perturbatively, and its nonperturbative evolution into a physical bound state. Despite decades of experimental and theoretical effort, a complete and consistent description of the charmonium production mechanism remains an open challenge \cite{Brambilla2011Review,Andronic:2015wma,Lansberg:2019adr}.

A commonly utilized framework to describe charmonium production is based on nonrelativistic QCD (NRQCD) factorization~\cite{Bodwin1995}. It expresses the charmonium production cross section as a sum over intermediate \(c\bar{c}\) Fock states:
\[ \mathrm{d}\sigma(H)=\sum_n \mathrm{d}\hat{\sigma}(c\bar c[n])\, \langle\mathcal O^H(n)\rangle , \]
where \(H\) denotes the observed charmonium state, and \(n\) specifies the quantum numbers of the intermediate charm-anticharm pair. \(\mathrm{d}\hat{\sigma}\) is the perturbatively calculable short-distance coefficient (SDC), and \(\langle \mathcal{O}^{H}(n)\rangle\) is the long-distance matrix element (LDME) encoding the nonperturbative transition from \(c\bar{c}[n]\) to \(H\). Because LDMEs cannot be computed from first principles, they are determined by fitting to experimental data. The framework gains predictive power through the universality assumption: LDMEs extracted from one process are applicable to others where NRQCD factorization holds~\cite{Bodwin1995,Butenschoen2011World,Ma2011NLO}.

Extraction of LDMEs based on the NRQCD framework has been carried out extensively for charmonium production~\cite{Ma:2010vd,Chao2012Polarization,Butenschoen2011World,Ma2011NLO,Shao2015Yields,Sun:2015pia}. In practice, however, LDMEs for different charmonium states are typically extracted independently: each state's parameters are fitted to its own experimental data, with contributions from other states either fixed from external inputs or treated as nuisance corrections. This sequential approach does not fully account for the correlations among states that arise through feed-down. For example, prompt \(J/\psi\) yields receive contributions from direct production as well as from decays of \(\psi(2S)\) and \(\chi_c\) states. The magnitude of these feed-down contributions depends simultaneously on the parent production cross sections, the branching fractions, the decay kinematics, and the experimental acceptance. When measurements of the heavier states are limited or imprecise, the resulting uncertainty propagates into the extracted direct-production LDMEs in a way that is difficult to track consistently. Previous analyses have addressed this partially by including feed-down corrections or by exploiting relationships among charmonium states to constrain parent production~\cite{Gong2013Prompt,Bodwin2016Fragmentation,Faccioli2020Global}, but the parameters of different states are rarely inferred within a single, coherent statistical framework.

In this work, we develop a joint Bayesian inference framework~\cite{Gelman2020Workflow} to simultaneously extract LDMEs for \(J/\psi\), \(\psi(2S)\), and \(\chi_c\), from measurements of prompt charmonium production cross sections and their ratios in proton-proton ($pp$) collisions at the Large Hadron Collider (LHC). Prompt production includes direct production and feed-down from heavier charmonium states, but excludes contributions from weak \(b\)-hadron decays. PYTHIA~8.317~\cite{Bierlich2022Pythia} is used to evaluate the leading-order (LO) SDCs, supplemented by parton showers, and to simulate the full feed-down structure. Incorporating measurements of all available charmonium states within a single inference preserves the correlations among production of different states through feed-down, maximizing their combined constraining power on a coherent set of LDMEs for different charmonia. This approach is especially valuable when direct measurements of higher charmonium states are limited. Based on the LDMEs extracted in this work, we further present predictions for the prompt \(J/\psi\) and \(\psi(2S)\) production cross sections in $pp$ collisions at a center-of-mass energy (\(\sqrt{s}\)) of 200 GeV, as accessible at the Relativistic Heavy Ion Collider (RHIC).

%% file: methodology.tex
\section{Methods}\label{sec:methods}

\subsection{Model construction}\label{sec:response}
Within the NRQCD framework, both \jpsi\ and \psip\ are produced through four channels: the color-singlet \({}^3S_1^{[1]}\) channel and the color-octet \({}^1S_0^{[8]}\), \({}^3S_1^{[8]}\), and \({}^3P_J^{[8]}\) channels. For the \(\chi_{cJ}\) family, heavy-quark spin symmetry relates the LDMEs of the three spin states, \(\langle\mathcal O^{\chi_{cJ}}(n)\rangle=(2J+1)\langle\mathcal O^{\chi_{c0}}(n)\rangle\), so the family is described by two shared LDMEs in the \(\chi_{c0}\) convention, \({}^3P_0^{[1]}\) and \({}^3S_1^{[8]}\). The remaining \(J\) dependence of the SDCs is taken from PYTHIA. The color-singlet LDME for \jpsi\ is fixed to \(\langle\mathcal O^{\jpsi}({}^3S_1^{[1]})\rangle=1.32\pm0.201\ \mathrm{GeV}^3\), following Ref.~\cite{Bodwin2008Singlet}, where the electromagnetic decay width is combined with a potential-model calculation to give a well-determined value. This leaves 9 LDMEs to be determined for the three charmonium states. All LDMEs of \(P\)-wave states are divided by \(m_c^2\), with \(m_c=1.4\ \mathrm{GeV}^{2}\), so that every LDME has units of \(\mathrm{GeV}^3\).

When evaluating the cross section of a given charmonium state, the excited states that feed down to it are included explicitly, each with its own LDMEs. For example, the prompt \jpsi\ cross section is
\begin{equation}\label{eq:feed-down}
\begin{aligned}
\mathrm d\sigma^{\jpsi}
={}&\sum_n \mathrm d\hat\sigma_n^{\jpsi}\,\me{\jpsi}{n}\\[2pt]
&+\mathcal B(\psip\to\jpsi+X)\sum_n \mathrm d\hat\sigma_n^{\psip}\,\me{\psip}{n}\\[2pt]
&+\sum_{J=0}^{2}\mathcal B(\chi_{cJ}\to\jpsi\,\gamma)\,(2J+1)
\sum_n \mathrm d\hat\sigma_n^{\chi_{cJ}}\,\me{\chi_{c0}}{n},
\end{aligned}
\end{equation}
where \(\mathcal B\) denotes a branching fraction and \(\me{\chi_{c0}}{n}\) is the LDME shared by the \(\chi_{cJ}\) family. Since \(\mathcal B(\psip\to\jpsi+X)\) is inclusive, the cascade \(\psip\to\chi_{cJ}\gamma\to\jpsi\gamma\gamma\) is contained in the second term. The third term therefore includes only directly produced \(\chi_{cJ}\).

To determine the SDCs and the feed-down structure, we generate about \(1.5\times10^9\) charmonium production events for each collision energy with PYTHIA~8.317~\cite{Bierlich2022Pythia}, using the settings: 
\begin{itemize}
\item \texttt{Charmonium:all = on}
\item \texttt{PhaseSpace:pTHatMin = 0.5} 
\item \texttt{PhaseSpace:pTHatMinDiverge = 0.5}
\item \texttt{SuppressSmallPT} user hook with parameters \((0.60, 4, \texttt{false})\). This hook damps the hard cross section at low transverse momentum (\(p_{\rm T}\)) and modifies the weighting of the strong coupling~\cite{Bierlich2022Pythia,Pythia8317Manual}.
\end{itemize}

The contribution of each NRQCD channel \(n\) is identified through its process code, and the SDCs are obtained by dividing out the LDME values used in PYTHIA. The event sample is large enough that the Monte Carlo (MC) statistical uncertainty in each bin is at most half of the corresponding experimental uncertainty, so MC fluctuations do not dominate.

For each accepted charmonium candidate, we follow its decay ancestry and assign it to the earliest physical charmonium state in the chain. The intermediate color-octet \(c\bar c[n^{(8)}]\) states are treated as part of the production of the meson itself, not as feed-down parents. For example, a \jpsi\ from a \psip\ decay contributes to the observed \jpsi\ spectrum, but its yield is governed by the \psip\ LDMEs. Assigning every contribution to its production origin in this way preserves the feed-down structure and allows the LDMEs of the whole charmonium family to be extracted simultaneously.

\subsection{Experimental data}\label{sec:data}

We use prompt charmonium measurements in $pp$ collisions at $\sqrt{s}=5.02$, 7, 8, and 13 TeV, as listed in Table~\ref{tab:data}, from ATLAS, CMS, LHCb, and ALICE experiments at the LHC. The selected data comprise 276 differential cross-section measurements: 179 for $J/\psi$, 89 for $\psi(2S)$, and 8 for $\chi_c$. An additional seven $\chi_{c2}/\chi_{c1}$ ratio measurements and three inclusive $\chi_c/J/\psi$ ratios bring the total to 286 observations. We require a minimum transverse momentum of $p_{\rm T}\geq10$~GeV for $J/\psi$ and $p_{\rm T}\geq11$~GeV for higher states~\cite{Bodwin2016Fragmentation}.

\begin{table}[!htb]
\centering
\small
\setlength{\tabcolsep}{3pt}
\renewcommand{\arraystretch}{1.02}
\caption{Prompt charmonium measurements in $pp$ collisions, comprising 286 data points across LHC experiments. Each row identifies the experiment, energy, observable and number of bins or data points retained for Bayesian inference.}\label{tab:data}
\begin{ruledtabular}
\begin{tabular}{lcl r}
Exp. & \shortstack{$\sqrt{s}$\\(TeV)} & Observable & Bins \\ \hline
ATLAS & 5.02 & $J/\psi$, $\psi(2S)$~\cite{ATLAS2018FiveTeV} & 33 \\
CMS & 5.02 & $J/\psi$~\cite{CMS2017FiveTeV} & 16 \\
ATLAS & 7, 8 & $J/\psi$ (8), $\psi(2S)$ (7, 8)~\cite{ATLAS2016SevenEight} & 135 \\
ATLAS & 13 & $J/\psi$, $\psi(2S)$~\cite{ATLAS2024Thirteen} & 40 \\
CMS & 7 & $J/\psi$, $\psi(2S)$~\cite{CMS2015Seven} & 30 \\
ALICE & 13 & $J/\psi$~\cite{ALICE2022Prompt} & 1 \\
LHCb & 7 & $J/\psi$~\cite{LHCb2011Jpsi} & 1 \\
LHCb & 13 & $J/\psi$~\cite{LHCb2015Jpsi} & 4 \\
LHCb & 7 & $\psi(2S)$ (2012 publication)~\cite{LHCb2012Psi2S} & 1 \\
LHCb & 7, 13 & $\psi(2S)$ (2020 publication)~\cite{LHCb2020Psi2S} & 7 \\
ATLAS & 7 & $\chi_{c1}$, $\chi_{c2}$~\cite{ATLAS2014Chic} & 8 \\
CMS & 7 & $\chi_{c2}/\chi_{c1}$~\cite{CMS2012ChicRatio} & 4 \\
LHCb & 7 & $\chi_{c2}/\chi_{c1}$~\cite{LHCb2013ChicRatio} & 3 \\
LHCb & 7 & Inclusive $\chi_c/J/\psi$ fraction~\cite{LHCb2012ChicFraction} & 3 \\
\hline Total & & & 286 \\ \end{tabular}
\end{ruledtabular}
\end{table}

\subsection{Bayesian determination of the LDMEs}
We use Bayesian inference~\cite{Gelman2020Workflow} to extract all 9 LDMEs simultaneously, together with their correlations. For a vector of LDMEs \(\boldsymbol{\theta}\) and a set of measurements \(\mathbf{y}\), the posterior probability density can be calculated as
\begin{equation}\label{eq:posterior}
p_{\mathrm{post}}(\boldsymbol{\theta}\mid\mathbf{y}) =\frac{\mathcal{L}(\mathbf{y}\mid\boldsymbol{\theta})\,p_{\mathrm{prior}}(\boldsymbol{\theta})}{Z}.
\end{equation}
The prior \(p_{\mathrm{prior}}\) encodes distributions assigned to LDMEs before the fit. The likelihood \(\mathcal{L}\) measures how well the predictions for a given \(\boldsymbol{\theta}\) agree with the measurements, and \(Z\) is a normalization constant independent of \(\boldsymbol{\theta}\). The posterior \(p_{\mathrm{post}}\) describes the LDME values inferred from the measurements.

The likelihood is
\begin{equation}\label{eq:likelihood}
\mathcal{L}\propto\exp\!\left(-\tfrac{1}{2}\chi^2_\mu\right),
\qquad
\chi^2_\mu=\sum_i\left(\frac{\mu_i^{\mathrm{th}}-y_i^{\mathrm{exp}}}{\delta_i}\right)^2,
\end{equation}
where \(y_i^{\mathrm{exp}}\) is the measured cross section or yield ratio in bin \(i\), \(\delta_i\) is its experimental uncertainty, obtained by adding the statistical and systematic uncertainties in quadrature, and \(\mu_i^{\mathrm{th}}\) is the corresponding prediction for a given set of LDMEs. Because covariance matrices are not available for most measurements, we treat the uncertainties as Gaussian and independent across bins. 

Taking the negative logarithm of Eq.~\eqref{eq:posterior} gives the loss function: 
\begin{equation}\label{eq:loss}
\mathcal{J}\equiv-\ln p_{\mathrm{post}}
=\tfrac{1}{2}\chi^2_\mu-\ln p_{\mathrm{prior}}+\mathrm{const}.
\end{equation}
We sample the posterior with the No-U-Turn Sampler (NUTS)~\cite{HoffmanGelman2014NUTS}, an adaptive Hamiltonian Monte Carlo algorithm. NUTS introduces an auxiliary momentum for each LDME and evolves the system along Hamiltonian trajectories driven by the gradient of \(\mathcal{J}\). The trajectory length is adapted automatically with a U-turn criterion, which allows efficient exploration of correlated directions in the nine-dimensional LDME space. The likelihood contribution to the gradient is
\begin{equation}\label{eq:gradient}
\frac{\partial(\chi^2_\mu/2)}{\partial\me{H}{n}}
=\sum_i\frac{\mu_i^{\mathrm{th}}-y_i^{\mathrm{exp}}}{\delta_i^2}\,
\frac{\partial\mu_i^{\mathrm{th}}}{\partial\me{H}{n}},
\end{equation}
where \(\partial\mu_i^{\mathrm{th}}/\partial\me{H}{n}\) is the response of bin \(i\) to channel \(n\) of state \(H\). The sampling is implemented in PyMC~\cite{AbrilPla2023PyMC}.

We assign independent Gaussian priors to the 9 LDMEs, with means and widths listed in Table~\ref{tab:priors}. The color-octet priors are centered at zero and are not truncated, so negative values are allowed. We adopt deliberately broad priors with widths about 20 times the PYTHIA default values, so that the extracted LDMEs are determined mainly by the experimental measurements rather than by the prior. The color-singlet LDMEs are restricted to positive values, consistent with their squared-amplitude form in the vacuum-saturation approximation~\cite{Bodwin1995}, and their priors are taken from external determinations. For \(\psi(2S)\), we use the color-singlet matrix element of \(0.274^{+0.042}_{-0.036}\,\mathrm{GeV}^{3}\) for \(\eta_c(2S)\) as reported in Ref.~\cite{Chung2008Singlet}. Multiplying by the spin-multiplicity factor \(2J+1=3\) gives the \(\psi(2S)\) production LDME, \(0.822^{+0.126}_{-0.108}\,\mathrm{GeV}^{3}\). For \(\chi_{c0}\), we use \(0.060^{+0.043}_{-0.029}\,\mathrm{GeV}^{5}\), obtained by combining the \(\chi_{c0}\) and \(\chi_{c2}\) two-photon decay determinations under heavy-quark spin symmetry~\cite{Chung2008Singlet}, and divide by \(m_c^2\) with \(m_c=1.4\)~GeV. Because the published uncertainties are asymmetric, we take the larger one as the Gaussian width, and the resulting Gaussian priors are truncated at zero. The \(J/\psi\) color-singlet LDME is fixed in the fit, and its uncertainty is included in the predictions.

\begin{table}[t]
\caption{Means and widths of the Gaussian LDME priors before truncation, in
\(\mathrm{GeV}^3\). \(P\)-wave LDMEs are divided by \(m_c^2\).}\label{tab:priors}
\begin{ruledtabular}
\begin{tabular}{llrr}
State $H$ & Channel $n$ & Mean & Width\\
\hline
$\jpsi$ & ${}^1S_0^{[8]}$ & 0 & 0.200\\
$\jpsi$ & ${}^3S_1^{[8]}$ & 0 & 0.238\\
$\jpsi$ & ${}^3P_0^{[8]}$ & 0 & 0.200\\
$\psip$ & ${}^3S_1^{[1]}$ & 0.822 & 0.126\\
$\psip$ & ${}^1S_0^{[8]}$ & 0 & 0.080\\
$\psip$ & ${}^3S_1^{[8]}$ & 0 & 0.100\\
$\psip$ & ${}^3P_0^{[8]}$ & 0 & 0.080\\
$\chi_{c0}$ & ${}^3P_0^{[1]}$ & 0.0306 & 0.0219\\
$\chi_{c0}$ & ${}^3S_1^{[8]}$ & 0 & 0.062
\end{tabular}
\end{ruledtabular}
\end{table}

We quote the posterior median of each LDME as its central value, with the central 68\% (16th--84th percentiles) and 95\% credible intervals as its uncertainties. Predictions are computed for every posterior sample, so the compensation between different LDMEs is propagated into all predicted observables.

%% file: results.tex
\section{Results and Discussion}
\label{sec:results}

\subsection{Description of the experimental data}

\begin{figure*}[!tp]
\centering
\includegraphics[width=\textwidth]
{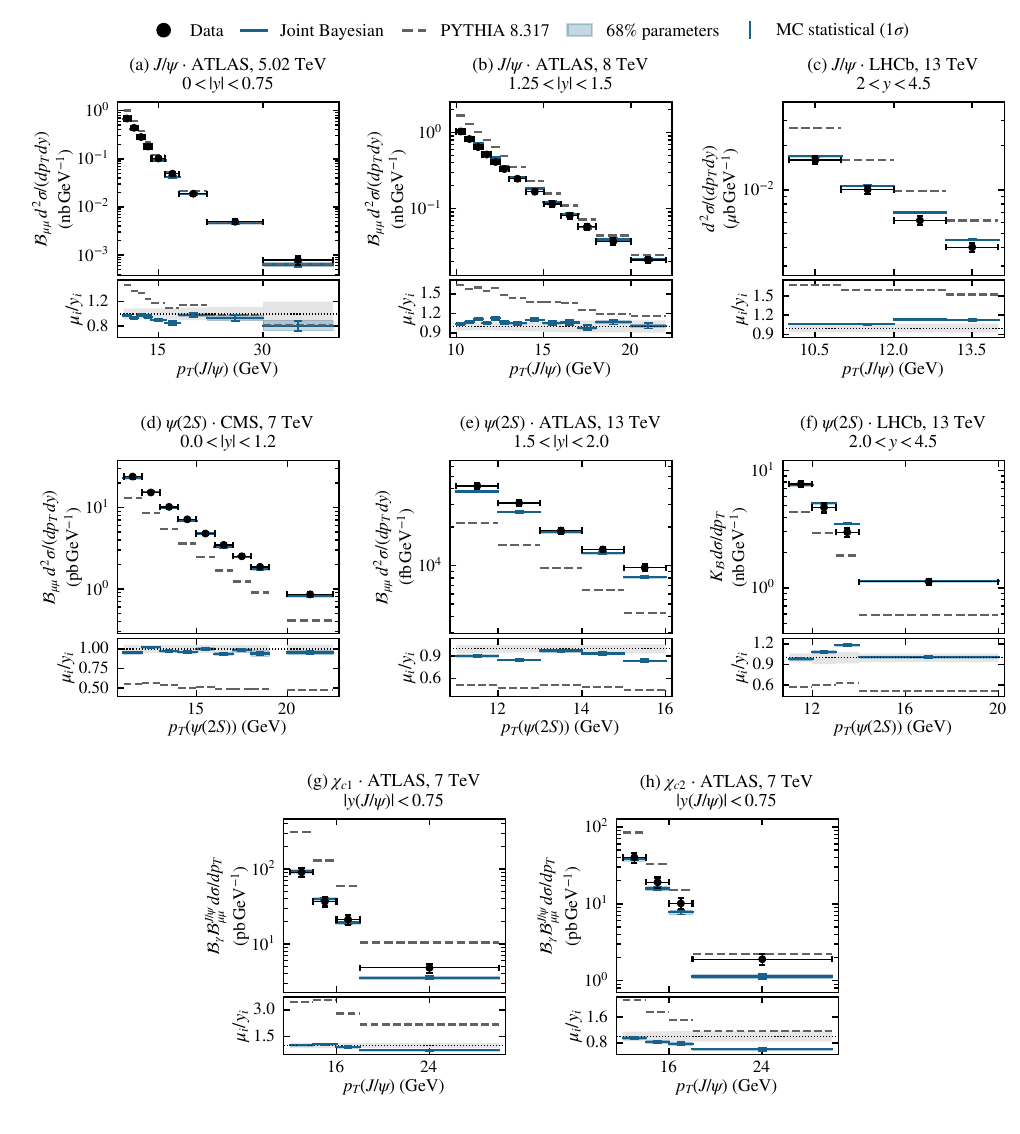}
\caption{Joint Bayesian description of the spectra for
(a--c) prompt \(J/\psi\), (d--f) prompt \(\psi(2S)\), and
(g,h) \(\chi_{c1,2}\), compared with experimental data
(black points). Solid blue lines show predictions using the
posterior-median LDME vector, blue shaded bands show the
central 68\% credible intervals from LDME parameter uncertainty,
and vertical blue bars indicate statistical errors from MC
simulations. Dashed gray lines show default PYTHIA~8.317.
Lower panels show model-to-data ratios with gray bands
indicating the relative experimental uncertainties.}
\label{fig:data-model-comp}
\end{figure*}

Figure~\ref{fig:data-model-comp} compares the posterior predictive distributions with selected \(J/\psi\), \(\psi(2S)\), and \(\chi_{c}\) measurements spanning mid- to forward rapidity at the LHC. Comparisons for all remaining cross-section spectra and ratios are given in Appendix~\ref{app:all_spectra}. The \(\chi_{c1}\) and \(\chi_{c2}\) states are reconstructed through their radiative decays \(\chi_{cJ}\to J/\psi\,\gamma\)~\cite{ATLAS2014Chic}, so their cross sections are presented as functions of the daughter \(J/\psi\) \(p_{\rm T}\). The joint Bayesian model describes the data well over several orders of magnitude in cross section: the model-to-data ratios scatter around unity, mostly within the experimental uncertainties. Mild tensions remain in a few places, most visibly in the \(\chi_{c2}\) spectrum and the highest-\(p_{\rm T}\) \(\chi_{c1}\) bin, where the model falls below the data by roughly 20--40\%. Default PYTHIA~8.317, by contrast, describes the data poorly. It overestimates prompt \(J/\psi\) production by 20--60\% and \(\chi_{c1}\) production by up to a factor of three, while underestimating \(\psi(2S)\) production by about a factor of two.

To provide a parameter set that can be used directly in PYTHIA, which accepts only non-negative LDMEs, we repeat the inference with all LDMEs restricted to non-negative values through positive-truncated priors. We label this variant ``Joint Bayesian~(+).''

\begin{table}[!htbp]
\centering
\caption{Comparison of default PYTHIA~8.317, Joint Bayesian,
and its positivity-constrained variant, Joint Bayesian~(+),
across the studied charmonium states. Models are evaluated
by \(\chi^2/\)ndf and by the 68\% and 95\% in-sample
predictive coverages.}
\label{tab:global-perform}
\begin{ruledtabular}
\begin{tabular}{lrrr}
Model & $\chi^2/$ndf & $C_{68}$ & $C_{95}$\\
\hline
PYTHIA 8.317       & 33.33 & N/A      & N/A\\
Joint Bayesian    & 1.18  & 67.83\%  & 93.71\%\\
Joint Bayesian (+)& 1.33  & 62.59\%  & 92.66\%\\
\end{tabular}
\end{ruledtabular}
\end{table}

To quantify the model performance, Table~\ref{tab:global-perform} lists \(\chi^2/\)ndf and the in-sample predictive coverages \(C_{68}\) and \(C_{95}\) for the two joint Bayesian fits and default PYTHIA~8.317. Here \(\chi^2\) is evaluated with the posterior-median LDMEs, and the number of degrees of freedom, ndf, is the number of observations (286) minus the number of free parameters (9). For PYTHIA, ndf = 286 as it does not contain free parameters. The joint Bayesian model yields \(\chi^2/\mathrm{ndf}=1.18\), a satisfactory description of the full dataset, while default PYTHIA~8.317, with \(\chi^2/\mathrm{ndf}=33.33\), is strongly disfavored. The coverages \(C_{68}\) and \(C_{95}\) are the fractions of measurements that fall within the central 68\% and 95\% posterior predictive intervals. For the joint Bayesian model, they are close to the nominal values, showing that the predictive uncertainties are well calibrated. These coverages are not defined for default PYTHIA, as its LDME uncertainties are not given. For the positivity-constrained fit, \(\chi^2/\)ndf rises from 1.18 to 1.33 and \(C_{68}\) drops. This moderate deterioration is the cost of excluding the negative-valued region preferred by the unconstrained fit. Nevertheless, the positivity-constrained fit still describes the data reasonably well.

\subsection{Posterior LDMEs and parameter correlations}

\begin{figure*}[!tp]
\centering
\includegraphics[width=0.82\textwidth]
{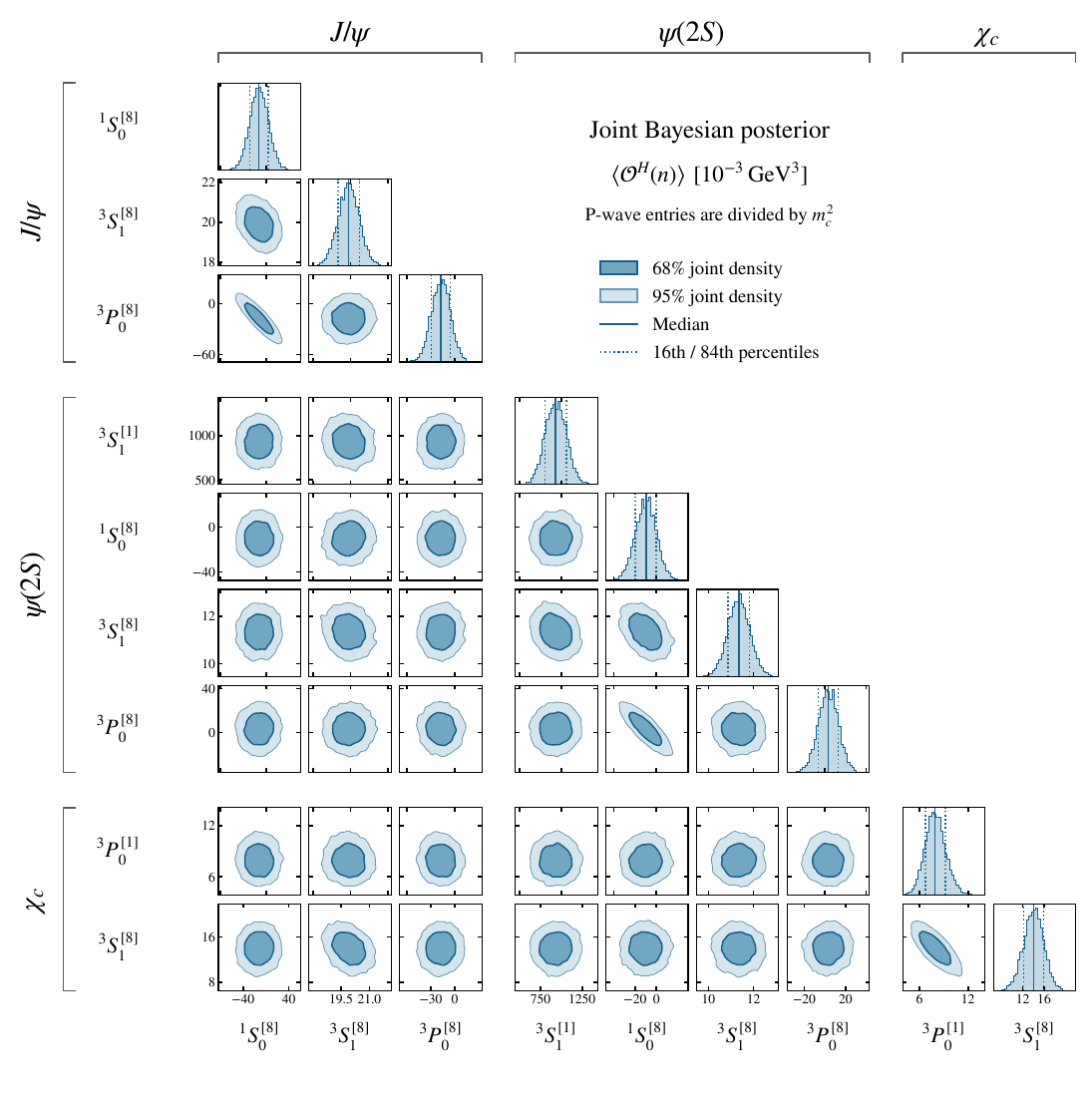}
\caption{Joint Bayesian posterior distributions for the
9 fitted LDMEs. Diagonal panels show marginal distributions,
where solid lines mark the medians and dotted lines the
16th and 84th percentiles. Off-diagonal panels show joint
distributions for each parameter pair, with dark and light
regions enclosing approximately 68\% and 95\% of the
posterior probability, respectively. Brackets group the
\(J/\psi\), \(\psi(2S)\), and \(\chi_{c}\) sectors.
Values are in \(10^{-3}\,\mathrm{GeV}^{3}\), with
\(P\)-wave entries divided by \(m_c^2\).}
\label{fig:LDME_Bayesian}
\end{figure*}

The joint Bayesian inference yields posterior distributions for the 9 free LDMEs of the charmonium family (the \(J/\psi\) \({}^3S_1^{[1]}\) LDME is fixed), as listed in Table~\ref{tab:ldmes_joint}. The marginal posteriors are shown in the diagonal panels of Fig.~\ref{fig:LDME_Bayesian}. All are unimodal and approximately Gaussian.

\begin{table}[!htbp]
\small
\centering
\renewcommand{\arraystretch}{1.22}
\setlength{\tabcolsep}{2.5pt}
\caption{LDME values for default PYTHIA~8.317 and the two
joint Bayesian fits, in \(10^{-3}\,\mathrm{GeV}^{3}\).
Listed values are posterior medians with central 68\%
credible intervals. \(P\)-wave entries are divided by
\(m_c^2\). The \(J/\psi\) singlet LDME and its uncertainty
are fixed according to Ref.~\cite{Bodwin2008Singlet}.
The \(\chi_{cJ}\) sector uses the \(\chi_{c0}\) values
listed here.}
\label{tab:ldmes_joint}
\begin{ruledtabular}
\begin{tabular}{lrrr}
Channel & PYTHIA & Joint & Joint\\
        & 8.317 & Bayesian & Bayesian~(+)\\
\hline
$J/\psi\ {}^3S_1^{[1]}$
 & 1160 & $1320\pm201$ & $1320\pm201$\\
$J/\psi\ {}^1S_0^{[8]}$
 & 10 & $-12\pm16$ & $0.69^{+1.10}_{-0.52}$\\
$J/\psi\ {}^3S_1^{[8]}$
 & 11.9 & $19.90\pm0.56$ & $17.58\pm0.40$\\
$J/\psi\ {}^3P_0^{[8]}/m_c^2$
 & 10 & $-17\pm12$ & $0.48^{+0.79}_{-0.36}$\\
\noalign{\vskip 3pt}
$\psi(2S)\ {}^3S_1^{[1]}$
 & 760 & $930\pm130$ & $1040\pm120$\\
$\psi(2S)\ {}^1S_0^{[8]}$
 & 4 & $-9.8^{+9.6}_{-10.1}$ & $1.13^{+1.67}_{-0.84}$\\
$\psi(2S)\ {}^3S_1^{[8]}$
 & 5 & $11.35\pm0.48$ & $10.41^{+0.21}_{-0.25}$\\
$\psi(2S)\ {}^3P_0^{[8]}/m_c^2$
 & 4 & $3.0^{+9.8}_{-9.7}$ & $1.21^{+1.85}_{-0.90}$\\
\noalign{\vskip 3pt}
$\chi_{c0}\ {}^3P_0^{[1]}/m_c^2$
 & 50 & $7.9^{+1.3}_{-1.2}$ & $6.1\pm1.1$\\
$\chi_{c0}\ {}^3S_1^{[8]}$
 & 3.1 & $14.1^{+1.8}_{-1.9}$ & $16.2\pm1.8$\\
\end{tabular}
\end{ruledtabular}
\end{table}

The off-diagonal panels of Fig.~\ref{fig:LDME_Bayesian} show the two-dimensional joint posteriors. Most parameter pairs are only weakly correlated. Three pairs, however, show strong anticorrelations: \({}^1S_0^{[8]}\) and \({}^3P_0^{[8]}\) for both \(J/\psi\) and \(\psi(2S)\), and \({}^3P_0^{[1]}\) and \({}^3S_1^{[8]}\) for \(\chi_c\). In each case, the two channels produce similar \(p_{\rm T}\) shapes in the fitted kinematic region, so their contributions can compensate each other and the data constrain mainly their combination~\cite{Cho1996Octet2,Ma2011NLO}. Consequently, the individual uncertainties on these LDMEs are considerably larger than the uncertainty on their combined contribution to the cross section.

Table~\ref{tab:ldmes_joint} lists the central values and 68\% confidence intervals of all LDMEs for both fits. The unconstrained fit prefers negative central values for the \(J/\psi\) \({}^1S_0^{[8]}\), \(J/\psi\) \({}^3P_0^{[8]}\), and \(\psi(2S)\) \({}^1S_0^{[8]}\) channels, although all three are within about \(1.5\sigma\) of zero. These are the channels involved in the strong \({}^1S_0^{[8]}\)--\({}^3P_0^{[8]}\) anticorrelation discussed above. Negative \(P\)-wave color-octet LDMEs are not unphysical in themselves and have been reported in previous next-to-leading-order (NLO) analyses~\cite{Butenschoen2011World,Gong2013Prompt}. The negative \({}^1S_0^{[8]}\) central values are likewise compatible with zero within uncertainties and reflect the strong anticorrelation with the corresponding \({}^3P_0^{[8]}\) channel, so that only their combination is well constrained by the data. When positivity is imposed, these channels become small and positive.

\FloatBarrier

\subsection{Comparison with previous extractions}

\begin{figure*}[!tp]
\centering
\includegraphics[width=0.90\textwidth]
{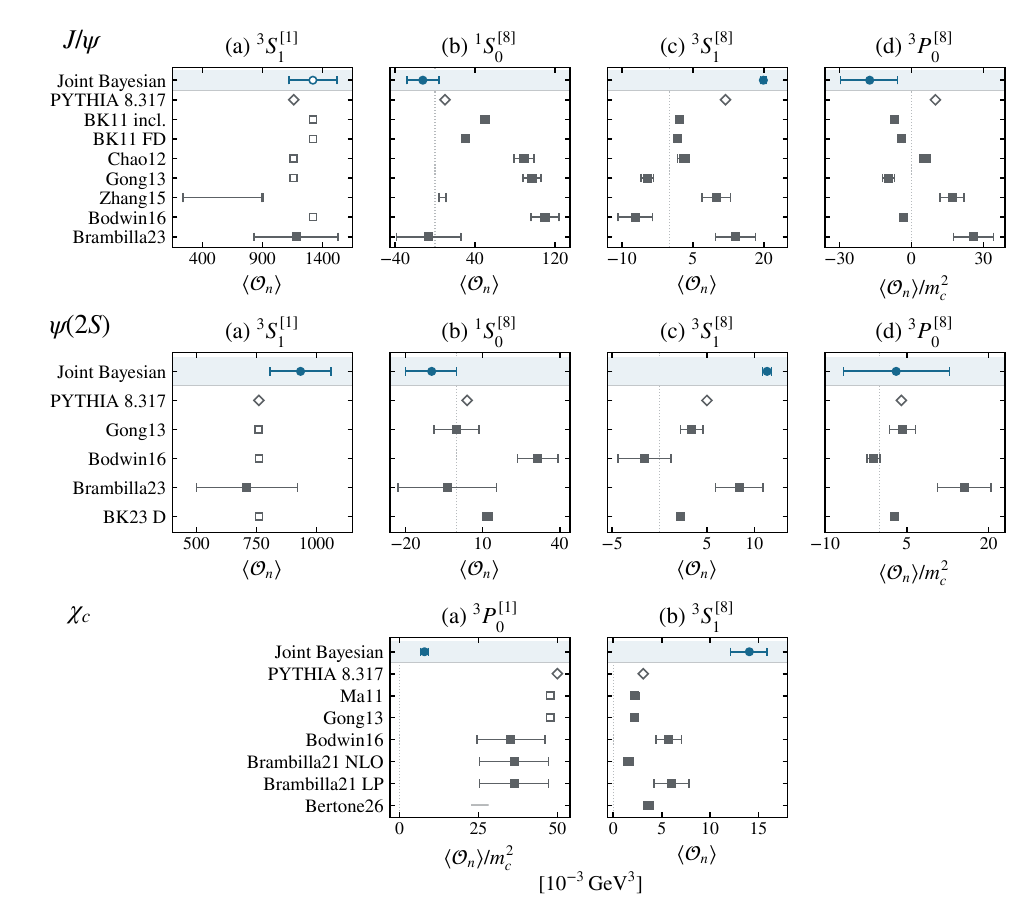}
\caption{Comparison of (top) \(J/\psi\), (middle)
\(\psi(2S)\), and (bottom) \(\chi_{c}\) LDMEs from the
joint Bayesian fit, default PYTHIA~8.317, and previous work:
BK11~\cite{Butenschoen2011World},
Chao12~\cite{Chao2012Polarization},
Gong13~\cite{Gong2013Prompt},
Zhang15~\cite{Zhang2015EtaC},
Bodwin16~\cite{Bodwin2016Fragmentation},
Brambilla23~\cite{Brambilla2023Inclusive},
BK23~\cite{Butenschoen2023Psi2S},
Ma11~\cite{Ma:2010vd},
Brambilla21~\cite{Brambilla2021Inclusive},
and Bertone26~\cite{Bertone2026Charmonia}.
Values are in \(10^{-3}\,\mathrm{GeV}^{3}\), with
\(P\)-wave entries divided by \(m_c^2\). Filled blue points
show posterior medians with central 68\% credible intervals,
while the open blue point shows the fixed singlet input
with its uncertainty. Gray symbols show previous results
with their published uncertainties, and open squares
indicate fixed inputs.}
\label{fig:ldme-comparison}
\end{figure*}

Figure~\ref{fig:ldme-comparison} compares the joint Bayesian LDMEs with those in default PYTHIA~8.317 and from previous extractions~\cite{Butenschoen2011World,Chao2012Polarization,Gong2013Prompt,Zhang2015EtaC,Bodwin2016Fragmentation,Brambilla2023Inclusive,Butenschoen2023Psi2S,Ma:2010vd,Brambilla2021Inclusive,Bertone2026Charmonia}. ``BK11 incl.'' and ``BK11 FD'' denote fits before and after subtracting estimated feed-down contributions. ``BK23 D'' represents the fit to unpolarized $\psi(2S)$ data with $p_{\rm T}>7$ GeV. ``Brambilla21 NLO'' and ``Brambilla21 LP'' refer to fixed-order NLO and leading-power-fragmentation improved NLO calculations, respectively. Our LDMEs are determined with PYTHIA's LO SDCs supplemented by parton showers.

For \(J/\psi\), previous extractions of the \({}^1S_0^{[8]}\) and \({}^3P_0^{[8]}\) LDMEs span a wide range, from negative values to \(\sim\!0.1\,\mathrm{GeV}^3\). Our values lie within this spread and are compatible with zero. The \({}^3S_1^{[8]}\) LDME, however, is larger than all previous determinations. Our fit therefore attributes the high-\(p_{\rm T}\) yield mainly to the \({}^3S_1^{[8]}\) channel, whereas several NLO analyses attribute it mainly to \({}^1S_0^{[8]}\)~\cite{Chao2012Polarization,Bodwin2016Fragmentation}. This could arise from differences in the extraction procedures, such as the order of SDCs, usage of different datasets, \(p_{\rm T}\) thresholds, and treatments of feed-down.

For \(\psi(2S)\), the color-singlet LDME is a free parameter in our fit, and its value is consistent within uncertainties with the potential-model value used in previous work and with the determination of Ref.~\cite{Brambilla2023Inclusive}. The color-octet pattern resembles that of \(J/\psi\): \({}^1S_0^{[8]}\) and \({}^3P_0^{[8]}\) are compatible with zero and with most previous results, while \({}^3S_1^{[8]}\) is larger.

The largest differences appear in the \(\chi_c\) sector. Our color-singlet LDME is about a factor of 4--6 smaller than previous values, many of which were fixed from potential models. Our color-octet LDME is correspondingly larger. As discussed above, these two parameters are strongly anticorrelated, and the available \(\chi_c\) data constrain their combined contribution much better than the individual ones.

\FloatBarrier

\subsection{Predictions for RHIC}

\begin{figure}[!htbp]
\centering
\includegraphics[width=0.8\columnwidth]
{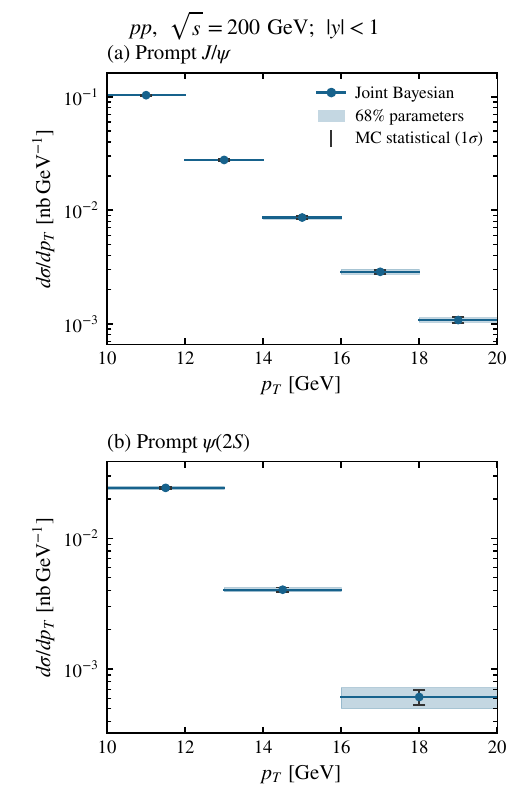}
\caption{Joint Bayesian predictions for the (a) prompt
\(J/\psi\) and (b) prompt \(\psi(2S)\) cross sections
within \(|y|<1\) in \(pp\) collisions at
\(\sqrt{s}=200\)~GeV. Points mark predictions using the
posterior-median LDME vector and horizontal lines indicate
the bin widths. Shaded bands show the central 68\% credible
intervals from LDME parameter uncertainty, and vertical
bars indicate statistical errors from MC simulations.}
\label{fig:200gev-pre}
\end{figure}

Using the joint posterior, we predict the prompt \(J/\psi\) and \(\psi(2S)\) \(p_{\rm T}\) spectra in \(pp\) collisions at \(\sqrt{s}=200\)~GeV for \(|y|<1\) and \(10\leq p_{\rm T}<20\)~GeV, as shown in Fig.~\ref{fig:200gev-pre}. This range lies in the high-\(p_{\rm T}\) regime used in the fit and within the acceptance of RHIC experiments.

Since no RHIC data enter the fit, these predictions provide a direct test of LDME universality across collision energies. Going from LHC to RHIC energy, charmonium measurements probe larger parton momentum fractions and different relative contributions of partonic subprocesses. Existing high-\(p_{\rm T}\) \(J/\psi\) measurements at RHIC~\cite{STAR2018Jpsi200,STAR:2026ljj} are inclusive and contain a nonprompt contribution from \(b\)-hadron decays, so we do not compare our prompt predictions with them. The sPHENIX experiment~\cite{PHENIX:2015siv,Campbell:2016cea}, whose silicon vertex detector provides precise vertex measurements, could measure prompt production with its recorded data.

\FloatBarrier

%% file: conclusion.tex
\section{Conclusions}\label{sec:conclusions}

We have developed a joint Bayesian framework for simultaneously extracting LDMEs for \(J/\psi\), \(\psi(2S)\), and \(\chi_c\), with the SDCs computed using PYTHIA~8.317. Applied to prompt charmonium measurements in \(pp\) collisions at the LHC, the extracted LDMEs describe the data well with a $\chi^{2}$/ndf of 1.18 on 286 data points. We also provide a positivity-constrained LDME set that can be used directly as PYTHIA input. Because these LDMEs are tied to PYTHIA's leading-order SDCs and parton shower, they should be used within the same configuration. We further present predictions for prompt \(J/\psi\) and \(\psi(2S)\) production in \(pp\) collisions at \(\sqrt{s}=200\)~GeV, which future analyses of RHIC data can test.

The Bayesian framework presented here is general and can be readily extended. It works on precomputed response matrices, so it can incorporate other theoretical inputs, such as NLO SDCs or other event generators, without changes to the inference itself. It can also include additional charmonium observables, including polarization, in-jet production, and event-activity dependence. Such extensions would provide better determinations of LDME values, more stringent tests of their universality, and a refined understanding of charmonium production.

%% file: backmatter.tex
\begin{acknowledgments}
The authors thank Geoffrey T. Bodwin of Argonne National Laboratory for discussions and feedback. The authors also greatly appreciate the opportunities and support provided by the High School Research Program at Brookhaven National Laboratory.
\end{acknowledgments}

\section*{Data and code availability}

Data and code are available from Yubo Liu upon request at \href{mailto:Boris_liu2008@outlook.com}{Boris\_liu2008@outlook.com}.

%% file: spectra_appendix/appendix_spectra.tex
\section{Additional spectrum and ratio comparisons}
\label{app:all_spectra}
\begin{figure}[!htbp]
\centering
\includegraphics[width=0.90\textwidth]{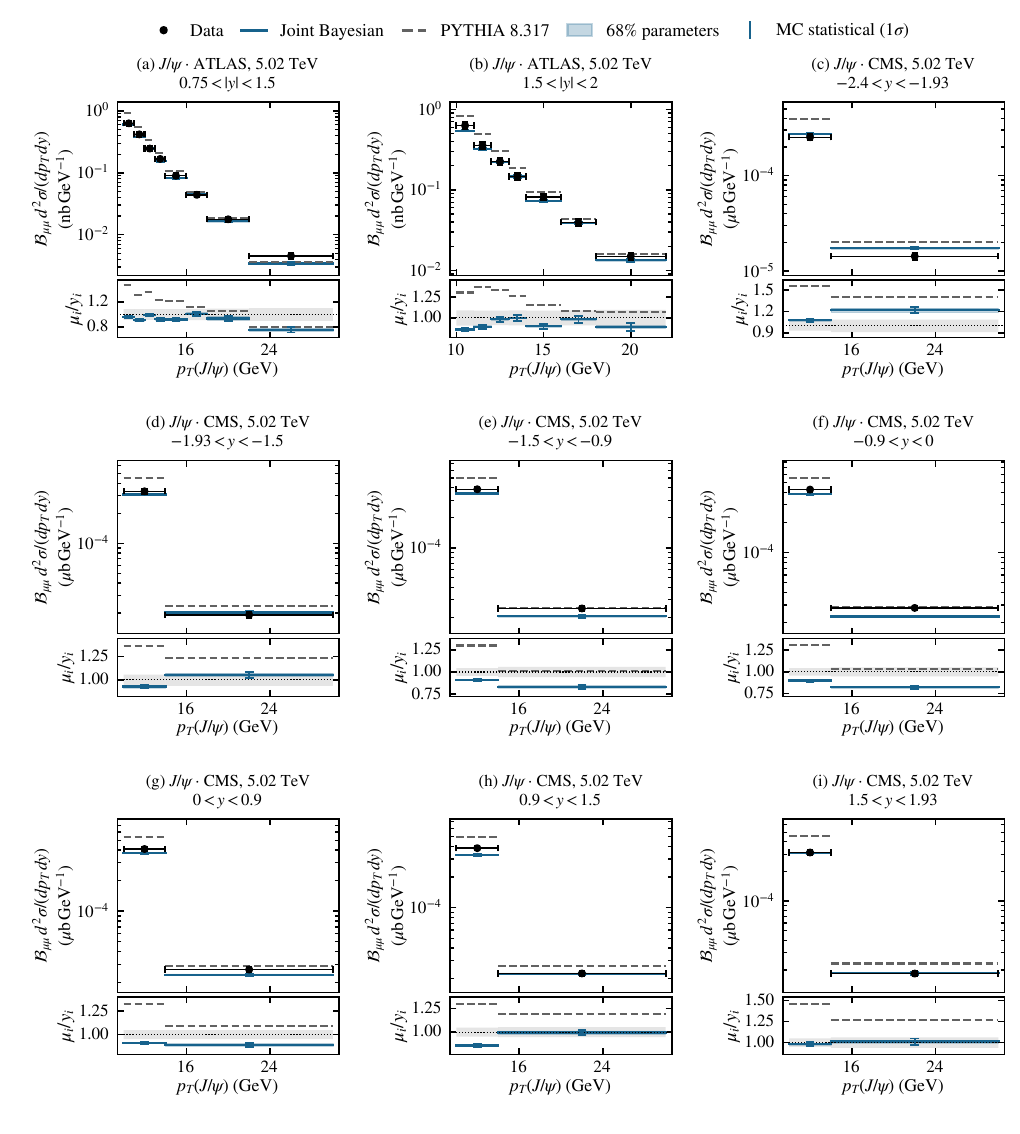}
\caption{Joint Bayesian description of the spectra for prompt \(J/\psi\) compared with experimental data (black points)~\cite{ATLAS2018FiveTeV,CMS2017FiveTeV}. Solid blue lines show predictions using the posterior-median LDME vector, blue shaded bands show the central 68\% credible intervals from LDME parameter uncertainty, and
vertical blue bars indicate statistical errors from MC simulations. Dashed gray lines show default PYTHIA~8.317. Lower panels show model-to-data ratios with gray bands indicating the relative
experimental uncertainties.}
\label{fig:additional_spectra_1}
\end{figure}

\begin{figure}[!htbp]
\centering
\includegraphics[width=0.90\textwidth]{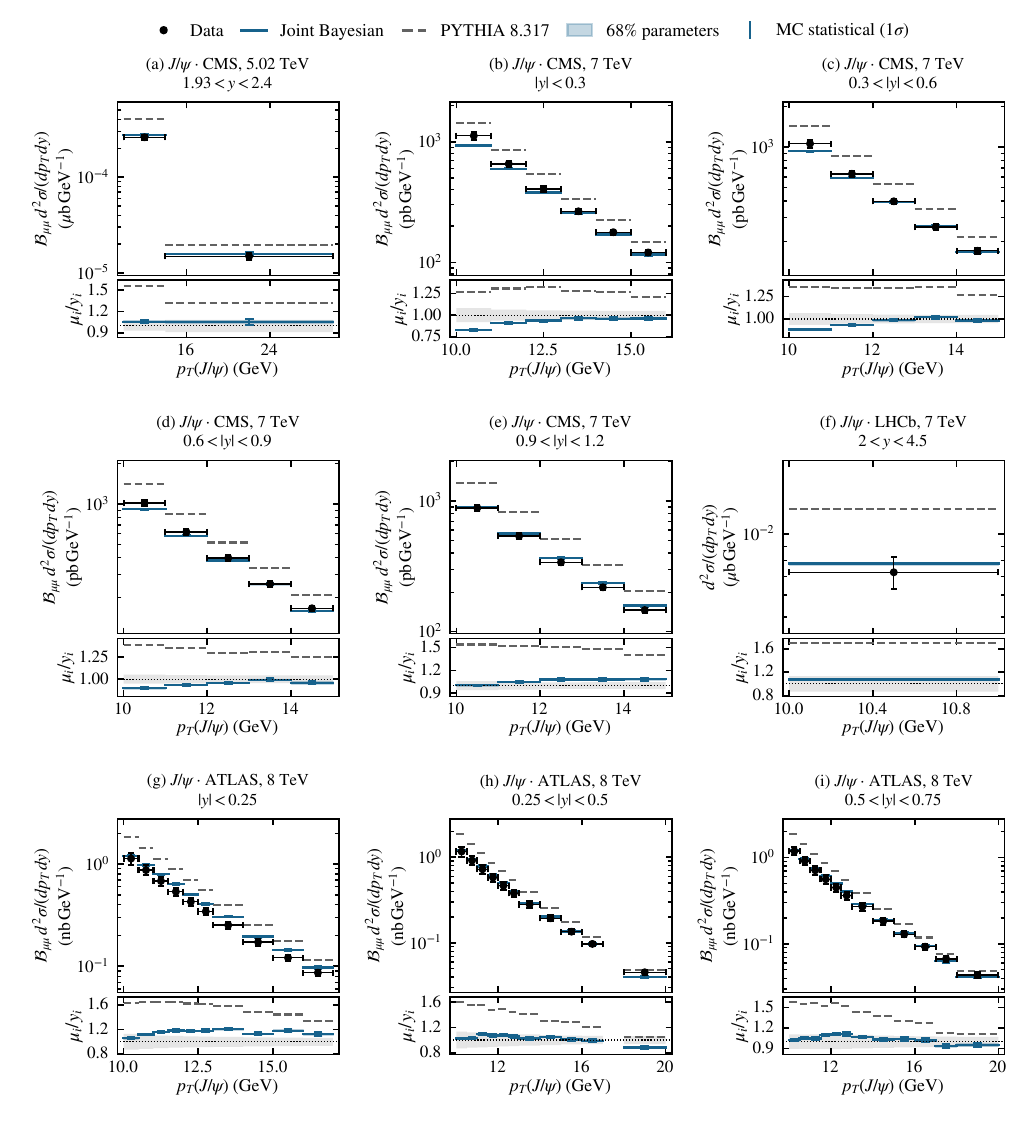}
\caption{Similar to Fig. \ref{fig:additional_spectra_1}, but for additional prompt $J/\psi$ data~\cite{CMS2017FiveTeV,CMS2015Seven,LHCb2011Jpsi,ATLAS2016SevenEight}.}
\label{fig:additional_spectra_2}
\end{figure}

\begin{figure}[!htbp]
\centering
\includegraphics[width=0.90\textwidth]{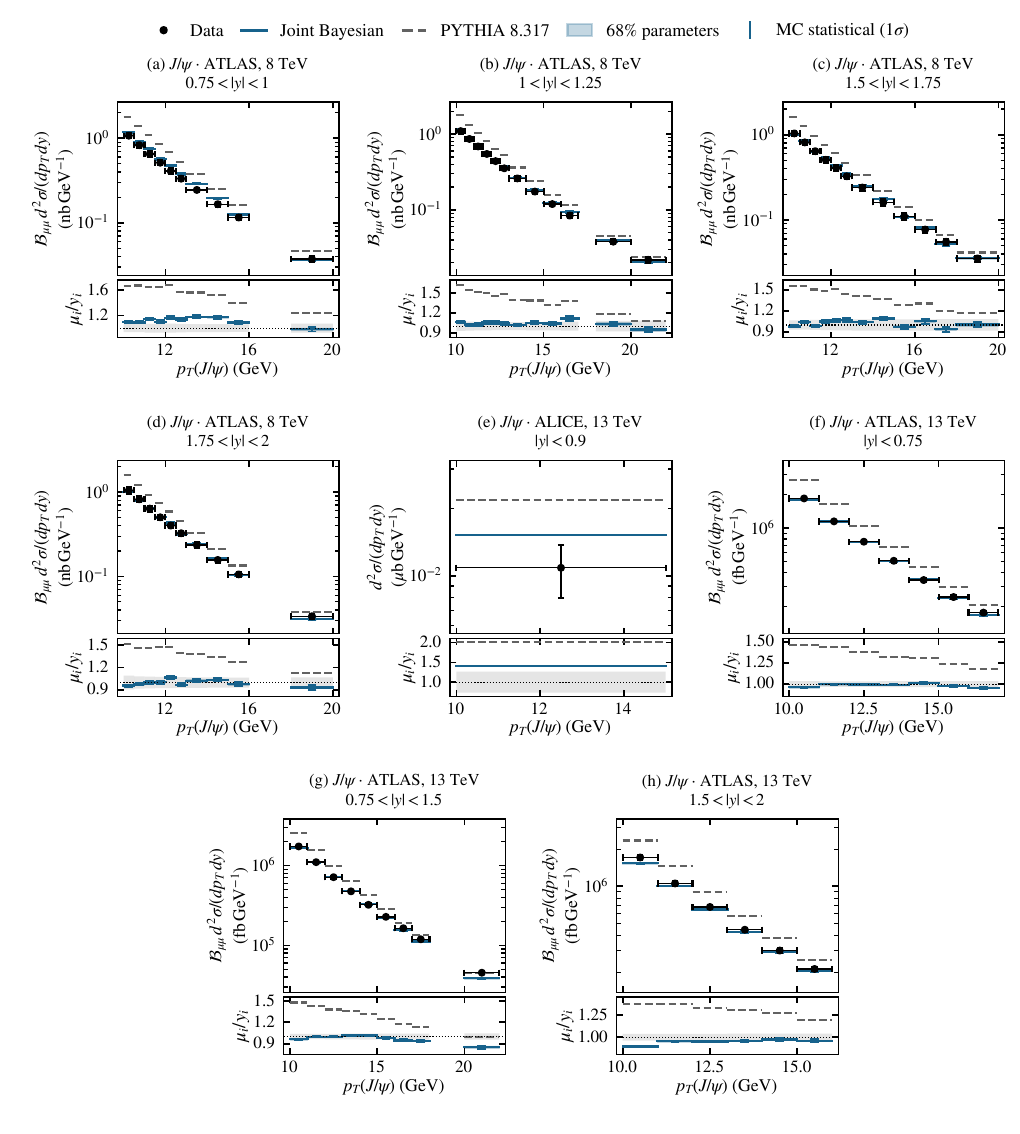}
\caption{Similar to Fig. \ref{fig:additional_spectra_1}, but for additional prompt $J/\psi$ data~\cite{ATLAS2016SevenEight,ALICE2022Prompt,ATLAS2024Thirteen}.}
\label{fig:additional_spectra_3}
\end{figure}

\begin{figure}[!htbp]
\centering
\includegraphics[width=0.90\textwidth]{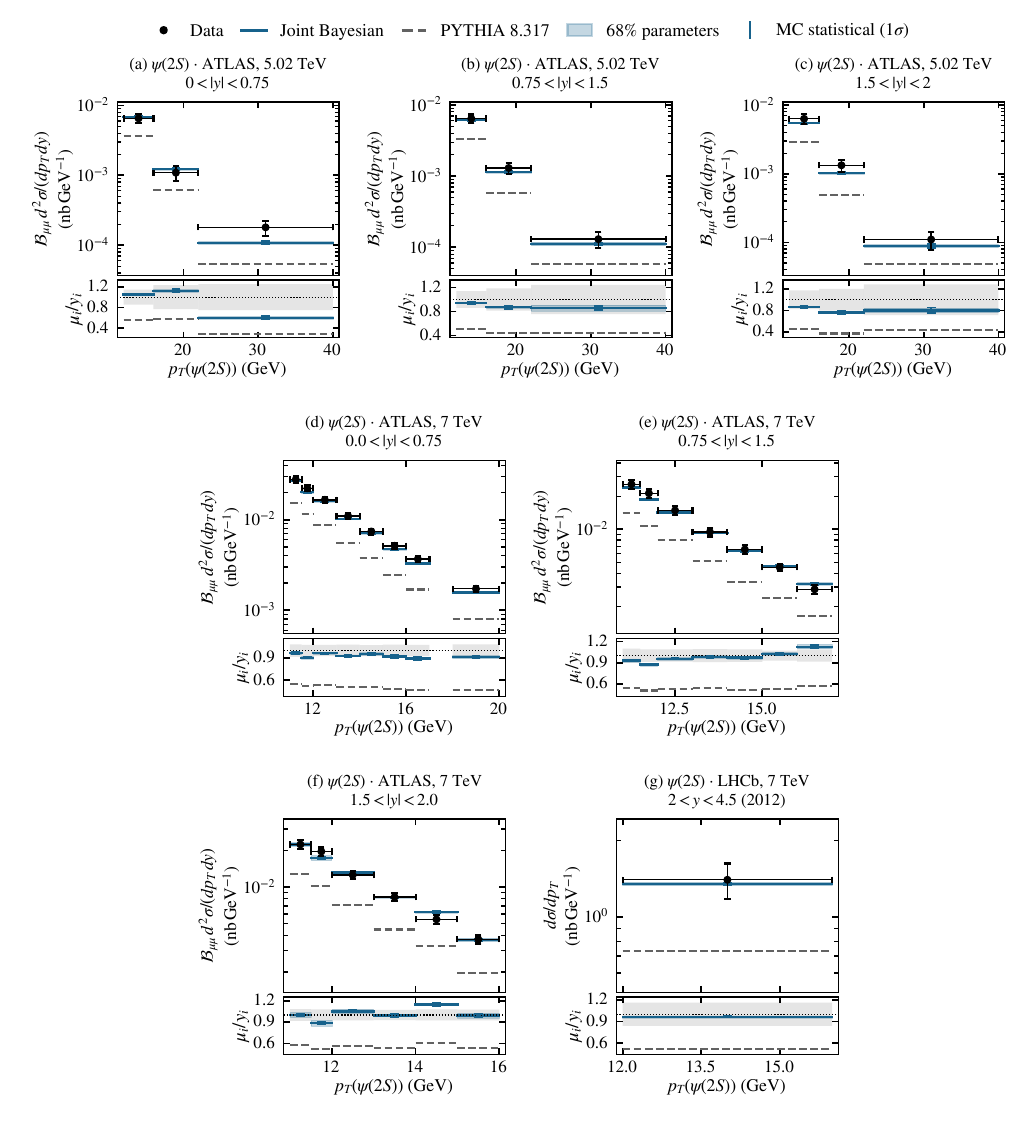}
\caption{Similar to Fig. \ref{fig:additional_spectra_1}, but for additional prompt $\psi(2S)$ data~\cite{ATLAS2018FiveTeV,ATLAS2016SevenEight,LHCb2012Psi2S}.}
\label{fig:additional_spectra_4}
\end{figure}

\begin{figure}[!htbp]
\centering
\includegraphics[width=0.90\textwidth]{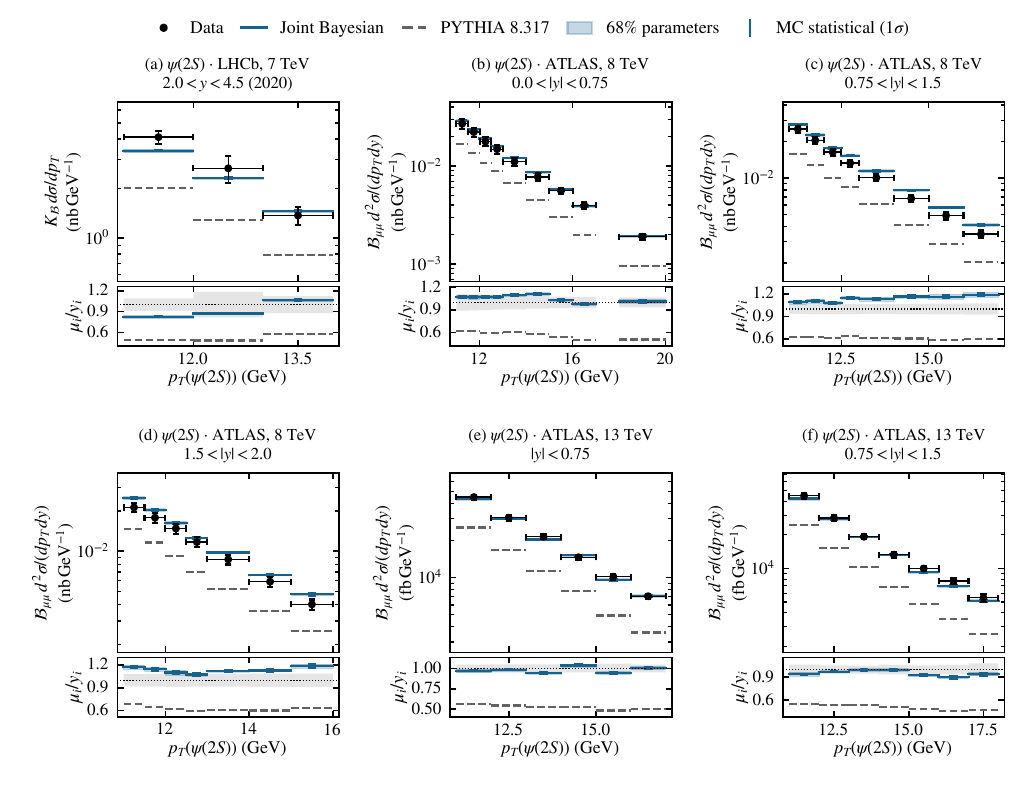}
\caption{Similar to Fig. \ref{fig:additional_spectra_1}, but for additional prompt $\psi(2S)$ data~\cite{LHCb2020Psi2S,ATLAS2016SevenEight,ATLAS2024Thirteen}.}
\label{fig:additional_spectra_5}
\end{figure}

\begin{figure}[!htbp]
\centering
\includegraphics[width=0.90\textwidth]
{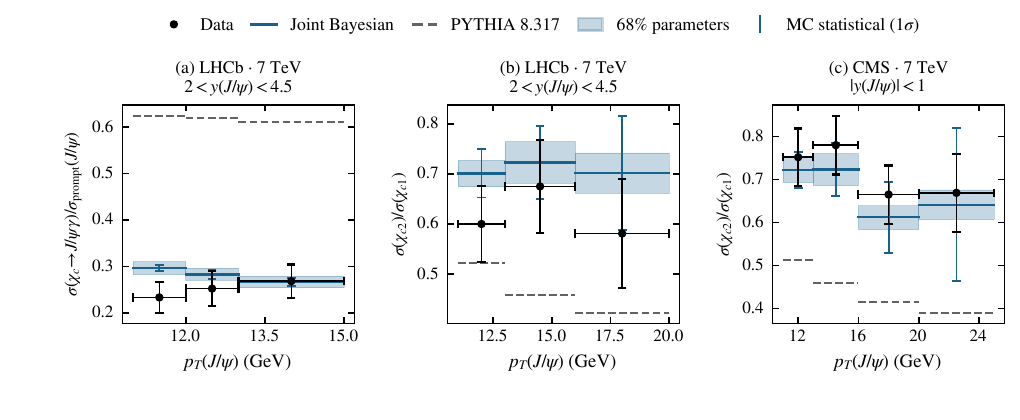}

\caption{Joint Bayesian description of (a) the LHCb
$\chi_c\to J/\psi\,\gamma$ feed-down fraction in prompt
$J/\psi$ production and the $\chi_{c2}/\chi_{c1}$ ratios
from (b) LHCb and (c) CMS at $\sqrt{s}=7$ TeV~
\cite{LHCb2012ChicFraction,LHCb2013ChicRatio,CMS2012ChicRatio}.
Model curves and uncertainties follow
Fig.~\ref{fig:additional_spectra_1}.
Black points show data with statistical, systematic, and
separately reported branching-fraction errors added in quadrature;
horizontal bars show bin widths.}

\label{fig:fitted-yield-ratios}
\end{figure}

This appendix presents the model-data comparisons not shown in
the main text. Figures~\ref{fig:additional_spectra_1}--\ref{fig:additional_spectra_5}
show 39 additional spectra: 26 $J/\psi$ spectra and
13 $\psi(2S)$ spectra, containing 224 fitted bins.
Figure~\ref{fig:fitted-yield-ratios} shows seven
$\chi_{c2}/\chi_{c1}$ production-ratio measurements
(three LHCb bins and four CMS bins) and three LHCb
measurements of the $\chi_c\to J/\psi\,\gamma$ feed-down
fraction in prompt $J/\psi$ production, all at
$\sqrt{s}=7$ TeV.
These figures cover the remaining 234 observations not displayed in the main text.
\clearpage

\clearpage

\clearpage

\clearpage

\clearpage

\clearpage
\clearpage